# Calculation of the $^{12}$C+$^{12}$C sub-barrier fusion cross section in an imaginary time-dependent mean field theory

A. Bonasera[1,2] and J,B, Natowitz[1]

1. Cyclotron Institute, Texas A&M University, College Station, TX 77843-USA.
2. Laboratori Nazionali del Sud-INFN, v. Santa Sofia 64, 95123 Catania, Italy.

## ABSTRACT

*The $^{12}$C+$^{12}$C sub-barrier fusion cross section is calculated within the framework of a Time Dependent Hartree-Fock (TDHF) based classical model using the Feynman Path Integral Method. The modified astrophysical S\*-factor is compared to direct and indirect experimental results. A good agreement with the direct data is found. In the lower energy region, where recent analyses of experimental data obtained with the Trojan Horse Method (THM) lead to contrasting results, the model predicts an S\* factor half way between those results. Low energy resonances revealed in the THM data are added to the calculation and the relative reaction rate in the Gamow region is calculated. The role of different resonances is discussed in detail and their influence on the reaction rate at temperatures relevant to stellar evolution is investigated.*

Careful measurements of the fusion cross-sections for $^{12}$C+$^{12}$C reactions are crucial to our understanding of massive stars and super-bursts from accreting neutron stars [1-4]. These measurements are especially compelling at energies close to or below the Gamow peak [4], i.e. below 2 MeV. Direct and indirect data measurements [5-13] exhibit many resonances, particularly notable when the data are expressed in terms of the modified astrophysical S* factor at center of mass energy $E_{cm}$ [14]:

$$S^*(E_{cm}) = E_{cm}\sigma(E_{cm}) \times \exp(87.12 E_{cm}^{-1/2} + 0.46 E_{cm}) = S(E_{cm}) \times \exp(0.46 E_{cm}) \qquad (1).$$

Efforts to make direct measurements at energies below 2.1 MeV are severely impacted by the rapidly diminishing reaction cross section at sub-barrier energies. To bypass the experimental difficulties of low fusion cross-sections at very low energies, indirect methods have been developed [15,16]. Recently, the $^{12}$C($^{14}$N, $^{2}$H)$^{24}$Mg indirect Trojan Horse Method, THM was employed to explore the lower energy $^{12}$C+$^{12}$C cross section [11]. These data revealed a wealth of resonances in S* at center of mass energies as low as 0.8MeV. It was concluded that the sub-barrier fusion cross section was much greater than previously estimated [17]. A critical re-analysis of that data, including Coulomb effects in the 3-body final channel and applying the Distorted Wave Born Approximation (DWBA) instead of the Plane Wave Approximation (PWA), resulted in much lower S* values by up to three orders of magnitude. These contrasting results have yet to be resolved.

Several macroscopic and microscopic models [17-24] offer a reasonable reproduction of the direct reaction data and they are sometime considered as determining an upper limit of the fusion cross-section [21,22]. Predictions of these models at energies below 2.1 MeV cannot reproduce the low energy THM results of reference [11] and are way above the values obtained in the Coulomb renormalization analysis of reference [12].

In reference [25], a microscopic model based on the Vlasov equation and the Feynman Path Integration Method, FPIM, was proposed. It succeeded in reproducing the then available direct data for the $^{12}C+^{12}C$ system rather well. The approach is also quite successful when applied to heavier nuclei [26] and spontaneous fission [27]. Before getting involved into complex numerical calculations, which may or may not be feasible at such low energies, it is instructive to simplify the heavy ion dynamics using a robust macroscopic model. In ref. [28-30], a Neck Model (NM), based upon the Time Dependent Hartree Fock, TDHF, approach was proposed and successfully reproduced fusion cross sections above the barrier [28] as well as deep inelastic [29] and fission [30] dynamics.

In the present work we extend the NM to sub barrier energies within the FPIM framework. Newton force equations are solved assuming the collective variables given by the center of mass distance **R** of the two colliding nuclei and their relative momenta **P**. The forces acting on the nuclei before the two nuclei touch, are given by the Coulomb and the nuclear Bass potentials [31]. After touching, the two nuclei are described as sections of spheres joined by a cylinder of radius $r_N$, the neck radius. In the rebounding phase, the nuclear geometry is given by two half spheres joined by sections of cones of radius $r_N$. Volume conservation is enforced which gives suitable relations between the neck radius and the relative distance of the two nuclei in good agreement with TDHF calculations [28]. In such a configuration the force is given by the surface tension times the perimeter of the neck, as for a liquid drop. During this stage nucleons are transferred through the neck, giving rise to one body-dissipation described by the Randrup-window formula [32].

In order to describe the sub-barrier fusion dynamics we go into imaginary times (FPIM) at the first (external) turning point [25-27] when the collective momentum $P=0$ fm$^{-1}$. In imaginary times, the second derivative of R with respect to (imaginary) time gets a $i^2=-1$ contribution and the Coulomb force becomes attractive while the nuclear part becomes repulsive. This accelerates the two nuclei towards each other until they come to a halt (because of the short range nuclear repulsion in imaginary times) at the second (internal) turning point. At this stage we switch from imaginary to real times again. If the nuclear force is strong enough the two nuclei fuse and this happens always for light nuclei like $^{12}C$. The probability of fusion for the lth-partial wave is given by [25]:

$$T_l = (1 + \exp\{2A\})^{-1} \qquad (2).$$

The action (in units of ℏ) is given by $A = \int_1^2 PdR$. The cross section is given by:

$$\sigma(E_{cm}) = \frac{\pi\hbar^2}{2\mu E_{cm}} \sum_{l=0}^{\infty} (2l+1) T_l P_l \qquad (3).$$

$P_l$ gives the probability of decay of the compound nucleus into different channels, μ is the nuclear reduced mass. Following [25], the present calculations have been performed only at zero impact parameter. In order to take into account the l-dependence of the transmission probability we shift the beam energy for each l as [33]:

$$T_l \approx T_0\left(E_{cm} - \frac{l(l+1)\hbar^2}{2\mu<R^2>}\right) \qquad (4),$$

μ<$R^2$> is an effective moment of inertia of two touching spheres at the internal turning point <R> and it is slowly varying with energy. Recall that for the symmetric

$^{12}$C+$^{12}$C system, because of angular momentum and parity conservation laws, only even l-values are allowed.

In order to take into account known resonances in the $^{24}$Mg compound nucleus, we increase the Bass potential strength at each resonance in the following way:

$$V_B \rightarrow V_B(1+g(x,\gamma,\sigma)).$$

Where $g(x,\gamma,\sigma) = \frac{1}{\sqrt{2\pi\sigma^2}} e^{-0.5\left(\frac{x-\gamma}{\sigma}\right)^2}$ is a Gaussian probability normalized to 1, x=$E_{cm}$/ΔE, γ=$E_{res}$/ΔE and σ=Γ/ΔE. $E_{res}$ is the energy and Γ is the natural width of the resonance taken from experimental data. ΔE is a characteristic energy for the model. Below the Coulomb barrier, $V_B$, we define it as ΔE=$V_{CB}$-$E_{CM}$. This is the energy violation during the time evolution in the classically forbidden region, i.e., during the imaginary time propagation.

The result of our calculation is presented in figure 1 where it is compared to experimental results. For the purpose of the calculation represented in figure 1 known values designated as 0$^+$ resonances in reference [11] are included. Although the 0.877 MeV resonance is designated 1$^-$ in the THM work [11] its population in the symmetric $^{12}$C+$^{12}$C reaction indicates that it is not [12]. It is probably 0$^+$ [13] and we include it as such. The resonance at 2.567 MeV, assumed to be 0$^+$ in reference [11] is not included. This is based upon the fact that there is no evidence for its population in either the direct reaction data or in the THM data. Our calculations would result in a huge peak if the resonance is treated as a 0$^+$ but this is in striking contrast to the direct data of ref. [8] and the more recent data of refs. [36,37]. The results of the R-Matrix analysis of reference [14] suggest that it is J$^\pi$≥ 3$^-$. Recall that the THM data is normalized in the 2.5 MeV region to the direct data [11] and there are no visible resonances at 2.567MeV. The resonance at 1.503MeV was assigned as a 2$^+$ in [11] but as a 0+ in [34]. If we treat it as a 0+ we observe a strong resonance, which seems to be confirmed by the data [11]. It disappears if it would be a 2$^+$ because the l=2 angular momentum gives no contribution at that energy. Higher angular momenta resonances are visible at 2.095MeV and 2.455MeV (2$^+$) c.m. energy. Higher angular momenta (l=4) and higher energy resonances result on the smooth increase with respect to the calculation with no resonances.

In reference [35] an analytical formula was derived for the astrophysical S-factor for Coulomb barrier penetration and a sharp cutoff at $R_N$=$r_1$+$r_2$ due to the nuclear force. Here the nuclei radii are $r_i$=1.4$A_i^{1/3}$. In the limit of zero beam energy we have [35]:

$$S_0 = S_G e^{\frac{4}{\hbar}\sqrt{2\mu Z_1 Z_2 e^2 R_N}} \quad (5).$$

$Z_i$ and e are the charge numbers and electric charge respectively. $S_G = \pi\hbar^2/(2\mu)$ is the astrophysical factor in the Gamow limit [35]. An extension to different l-values is possible, here we consider the dominant l=0 channel only [35]. The relation between S and S* discussed above, eq. (1), gives a smooth energy dependence as shown in figure 1. For our system $S_0$=10$^{16}$MeVb, in fair agreement with the model calculations absent resonances.

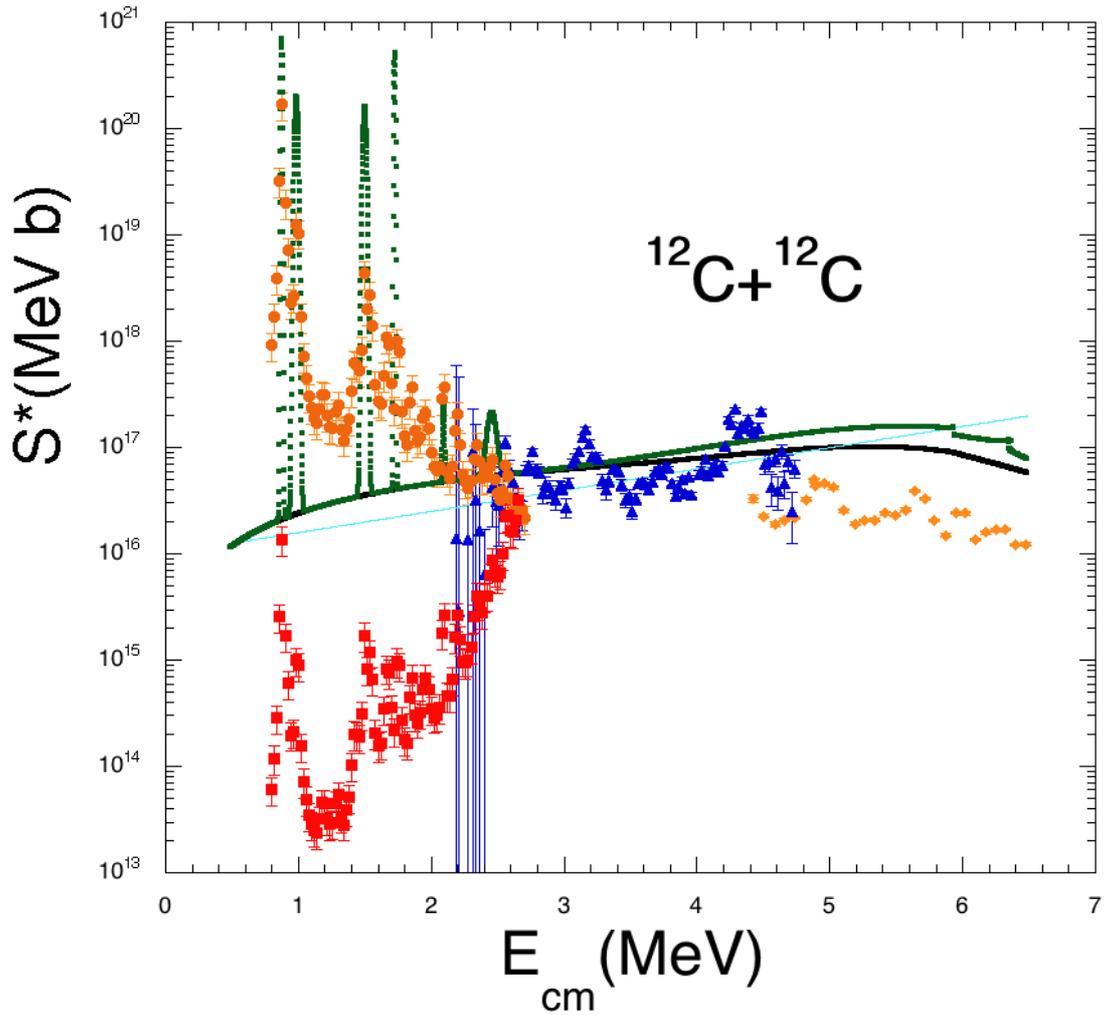

Figure 1. Astrophysical S* factor as a function of the center of mass energy in $^{12}C + ^{12}C$ collisions. The calculated smooth black curve increasing from $10^{16}$ MeVb to $\sim 10^{17}$ MeVb is obtained when no resonances are included. The analytical result, eqs.(1,5), is given by the cyan curve [35]. The result including resonances with their natural widths is represented by small squares (green on line). Experimentally derived data are orange diamonds [7], blue triangles [8], orange circles [11] and red squares [12]. The latter two represent conflicting THM analyses of the same THM data [11,12]. Notice that the experimental resolution of 30 KeV, not included in the calculations, broadens the peaks. The double narrow peaks below 1MeV are included in the calculations assuming they are both $J^\pi=0^+$ [11]. The resonance at 2.56MeV [11] is not included since it is most probably $J^\pi > 3^-$ [13]. For more recent direct data see references [36-37].

As seen in this figure, the two different proposed analyses of the THM reaction data lead to S* values well above or well below those of the present calculation without resonances or the simple analytical formula [35]. These discrepancies need to be resolved. This might be possible using data obtained with the THM approach but for a system, which does not present the same Coulomb problem for one of the outgoing particles. A suitable reaction could be $^{13}C(^{12}C,n)\ldots$where the neutron may act as a spectator [12]. Our calculations including the natural resonances are closer to the results of ref. [11] without Coulomb renormalization.

In figure 2 we present *relative* calculated reaction rates in the Gamow peak region vs $T^9$. For this figure the calculated reaction rate was divided by a reference one obtained using a constant $S_0= 1 \times 10^{16}$ MeVb. The green circles: ratio without resonances; red squares: ratio with all included resonances, see figure 1.

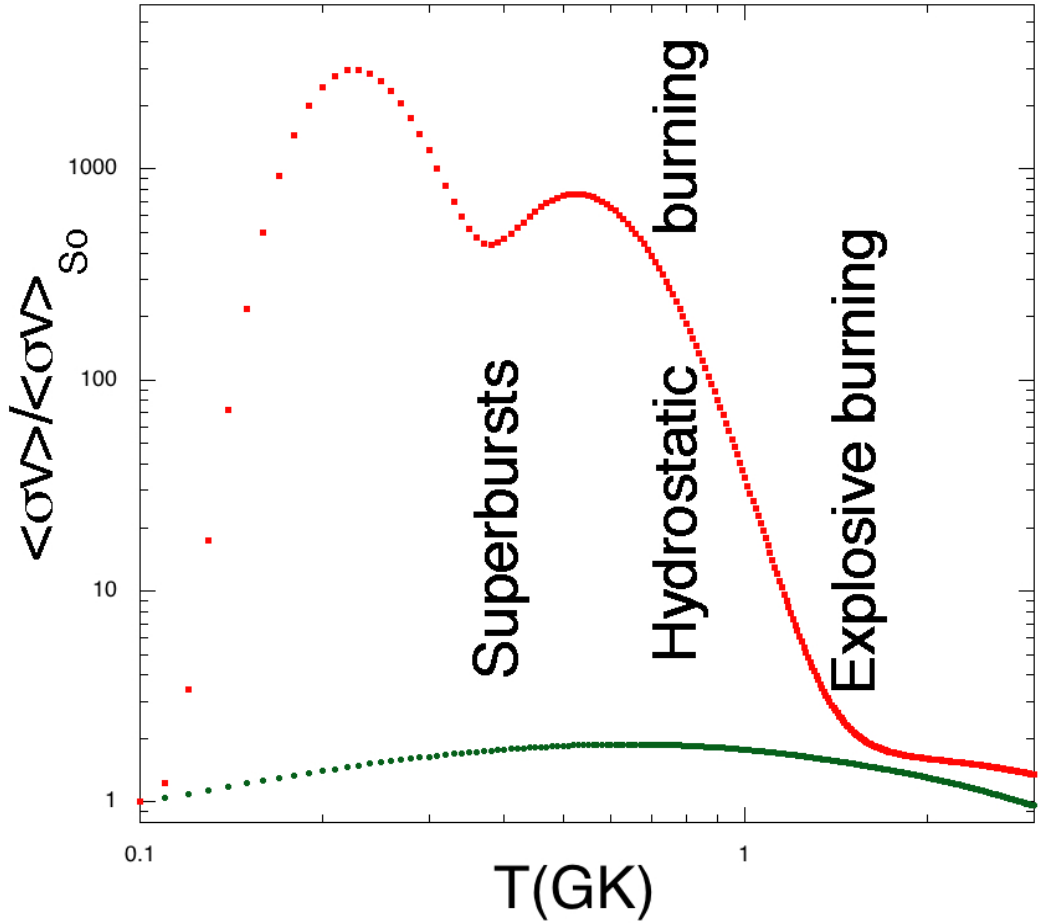

*Figure 2. Ratio of reaction rate divided by assumed constant $S_0=2 \times 10^{16}$ MeVb vs $T^9$. Green circles: ratio without resonances; red squares: ratio with all included resonances. Some typical regions for carbon burning are indicated in the figure, compare to ref. [11].*

We have excluded the 2.567 MeV resonance, the inclusion of which would result in more than one order of magnitude increase in the region of explosive burning, T>1GK. In reference [13], the resonance is discussed in great detail and it is excluded as $0^+$ in contrast to reference [11], see also figure 1. Also relevant is the role of the 0.877 MeV resonance. If this is a $1^-$ resonance as claimed in ref. [11] then it should not be included in the calculations because of spin and parity violation. The reaction rates would hugely decrease at the lowest temperatures, but it is most probably a $0^+$ and its inclusion results in the large ratio at low $T_9$. This is crucial to understand the early evolution of carbon stars. Compare the red squares to the results displayed in figure 3 of reference [11]. Some difference may come from the resonance widths increased by the finite detector resolution.

In conclusion, in this paper we have introduced a TDHF inspired macroscopic model and extended it to sub-barrier energies using the Feynman Path Integral Method i.e. solving Newton dynamics below the barrier in imaginary times. The model has no free parameters and reproduces reasonably well the direct data for the $^{12}C+^{12}C$ reaction. Extending the model by including low energy resonances in the region below that of the direct data we have calculated the Astrophysical S* factor for

$^{12}$C+$^{12}$C in the range of $E_{CM}$ =0.4 to 6 MeV. The reported analyses of the THM indirect data for $^{12}$C($^{14}$N, $^{2}$H)$^{24}$Mg lead to values which are in some agreement to our model results [11] but in contrast to [12] with the Coulomb corrections. It is important to stress that, in experiments, the widths of the narrow resonances would be increased by the finite detector resolution. The results depend crucially on the spin and parity assigned to each resonance in the region of interest. While we are quite confident that the 2.567MeV resonance is NOT a $0^+$ resonance, the 0.877MeV resonance is included as such. The reaction rates and thus the evolution of carbon rich stars depend crucially on the measured and calculated astrophysical factor as of figure 1. We believe that our simple model greatly helps to clarify some aspects and calls for a better determination of resonances in the energy of interest and especially their spin and parity: a challenge for future experiments.

ACKNOWLEDGEMENTS

We thank Profs. A. Tumino and A. Zhanov for discussions and providing the THM data. This work was supported in part by the United States Department of Energy under Grant # DE-FG03-93ER40773.